\documentclass[pra,showpacs,twocolumn]{revtex4-1}

\usepackage[utf8x]{inputenc}
\usepackage{amsmath,amssymb,amsfonts,subfigure,braket,color,dsfont}
\usepackage[english]{babel}
\usepackage[dvips]{graphicx}
\usepackage{sistyle}

\usepackage{subfig}

\newcommand{\norm}[1]{\left\| #1 \right \|}

\renewcommand{\H}{\mathcal{H}}

\DeclareMathOperator{\tp}{\otimes}
\DeclareMathOperator{\trace}{Tr}
\DeclareMathOperator{\Var}{Var}

\newcommand{\abs}[1]{\left| #1 \right|}

\newcommand{\id}{\mathds{1}}

\newcommand{\info}[1]{\textcolor{blue}{(#1)}}

\renewcommand{\info}[1]{}

\begin{document}
\title{Interacting spins in a cavity: finite size effects and symmetry-breaking dynamics}

\author{S{\o}ren Gammelmark}
\author{Klaus M{\o}lmer}
\affiliation{Lundbeck Foundation Theoretical Center for
  Quantum System Research, Department of Physics and Astronomy,
  University of Aarhus, DK 8000 Aarhus C, Denmark}

\begin{abstract}
We calculate the ground state and simulate the dynamics of a finite chain of spins with Ising nearest-neighbor interactions and a Dicke collective spin interaction with a single mode cavity field.
We recover the signatures of first and second order phase transitions predicted by mean field theory, while for small chains, we find significant and non-trivial finite size effects.
Below the first order phase transition, even quite large spin chains of 30-40 spins give rise to a mean photon number and number fluctuations significantly above the mean field vacuum result.
Near the second order phase critical point, our calculations reveal photon number fluctuations that grow beyond Poisson statistics with the size of the spin chain.
We simulate the stochastic evolution of the system when the cavity output field is subject to homodyne detection.
For an initial state close to the first order phase-transition the random character of the measurement process causes a measurement-induced symmetry-breaking in the system.
This symmetry-breaking occurs on the time-scale needed for an observer to gather sufficient information to distinguish between the two possible (mean-field) symmetry-broken states.
\end{abstract}

\pacs{03.65.Yz, 03.65.Ta, 02.70.-c, 75.10.Pq, 05.70.Fh, 05.50.+q}

\maketitle

\section{Introduction}

The problem of a single quantum system coupled to a large number of mutually interacting systems is found in a wide range of physics problems: an electron spin in a quantum dot or in an NV-center interacting with nuclear spins, an atom interacting with a continuum of quantized field modes, a cavity field mode interacting with an ensemble of absorbing atoms.

In this manuscript we present a matrix product state (MPS) analysis of the quantum states of a many body system, combining a one dimensional chain of spin-1/2 particles with nearest neighbor interaction and a cavity oscillator mode with a quantized radiation field which is linearly coupled to all the spins.
These two interactions are separately known as the Ising and the Dicke models with known many-body properties and phase diagrams, and  models which combine the spin-spin and spin-oscillator models interactions have been studied by mean field theory in recent publications \cite{lee_first-order_2004,gammelmark_phase_2011-1,tian_circuit_2010}. By use of the matrix product state analysis we will be able to  study non-mean field and finite size effects in such highly relevant theoretical models.

Numerous examples of quantum systems with local interaction in one spatial dimension have revealed a decrease of entanglement with the distance between pairs of particles.
Inspired to a large extent by insights from quantum information science, matrix product states have been developed as a very powerful method for the representation and solution of quantum many-body problems in physics.
The basic idea is to explicitly use the restriction of the entanglement between remote components of the full quantum system in a suitable variational ansatz for the quantum state.
A review of the development of these ideas, examples of their application to different systems, and their equivalence to the so-called density matrix renormalization group (DMRG) is offered in \cite{verstraete_matrix_2008,schollwock_density-matrix_2005}.
The interaction of the the cavity field with all the other subsystem degrees of freedom seemingly violates the usual requirement for matrix product state to apply.
We show, however, that by an adequate representation we can indeed implement and validate the method even with a globally coupled mode.

The paper is organized as follows.
In Sec. II, we recall the basic idea of the matrix product state method, and we describe how one may include a non-local quantum degree of freedom, and why this does not violate essential properties of the method.
In Sec. III, we introduce the Dicke-Ising Hamiltonian, and we present our numerical results for the ground state of the system as a function of the interaction parameters.
In symmetry breaking mean field theory, first and second order phase transitions have been identified in the model \cite{lee_first-order_2004,gammelmark_phase_2011-1}, and we show how these transitions reveal themselves in systems of finite size without breaking of the symmetries of the Hamiltonian.
Finally, in Sec. IV, we show how measurements on the oscillator mode can be simulated within the matrix product state method, and how the random measurement outcomes in a natural manner drive the system into symmetry broken states.
In Sec. V, we summarize and conclude.

\section{The Matrix Product State method}

For one-dimensional lattice systems, matrix product states (MPS) \cite{vidal_efficient_2004,schollwock_density-matrix_2005,vidal_class_2008,verstraete_matrix_2008} provide an efficient parametrization of the many-body Hilbert space.
For a system with $L$ sites and a conventional tensor product basis $\ket{i_1}\tp\ket{i_2}\tp\ldots\ket{i_L} \equiv \ket{i_1\ldots i_L}$ the matrix product state parametrization of the expansion coefficients takes the form
\begin{align}
 \ket{\psi} = \sum_{i_1\ldots i_L} \trace\left( \prod_{j=1}^L A^{[j],i_j} \right)\ket{i_1,\ldots i_L}. \label{eq:MPSansatz}
\end{align}
For each basis state $|i_j\rangle$ on a site $j$ one defines a $D\times D$ matrix $A^{[j],i_j}$, or, equivalently, for each site $j$ one defines a $D\times D\times d$ tensor, $A^{[j]}$, where $d$ is the single-site Hilbert space dimension.
By multiplying these matrices together and extracting the trace as specified in (\ref{eq:MPSansatz}), we thus obtain a set of expansion coefficients on the tensor product basis states $\ket{i_1\ldots i_L}$.
The total number of parameters in all matrices equals $L D^2 d$ and grows only linearly with the number of sites, unlike the number of tensor product basis states which grows as $d^L$.
The validity of the expansion thus relies strongly on the restriction of the quantum system to a small part of its Hilbert space caused, in parts, by the locality of interactions.

For $D = 1$ all expansions coefficients factor into products of single site amplitudes, $A^{[j],i_j}$, and the state in (\ref{eq:MPSansatz}) is simply a product state, valid for non-interacting and non-entangled particles.
For higher values of the matrix dimension $D$, however, the parametrization (\ref{eq:MPSansatz}) forms the basis for a variational calculation of the system properties, which supports an increasing amount of entanglement between the subsystems.
There is a host of theoretical work regarding which states matrix product states can represent efficiently (i.e. with a reasonably low $D$) \cite{schuch_peps_2010,verstraete_criticality_2006,perez-garcia_matrix_2006,PhysRevLett.94.140601,aguado_topology_2007}.
In general the results reveal that one-dimensional systems with nearest-neighbor interactions and, in particular, gapped one-dimensional systems can be represented efficiently.
It is also possible, however, to efficiently describe systems with interactions  with both exponential and polynomial dependence on the distance between the particles \cite{hauke_complete_2010,pirvu_matrix_2010}.
A key advantage of matrix product states is the ability to evaluate the energy variance $\braket{\psi|H^2|\psi} - \braket{\psi|H|\psi}^2$ which gives an absolute bound on the accuracy of the state-energy, and also a very reliable indicator for convergence even for non-standard problems.

In this paper we investigate a more general system than a one-dimensional chain.
We will introduce one more degree of freedom which is allowed to interact simultaneously with all the sites of the chain.
This situation occurs in e.g. the Dicke-model \cite{dicke_coherence_1954,hepp_equilibrium_1973,gammelmark_phase_2011-1} and Spin-star models \cite{bortz_exact_2007,hamdouni_exact_2006,hutton_mediated_2004,bortz_dynamics_2010}.
The added degree of freedom mediates a coupling and hence permits correlations and entanglement between remote spins which is independent of their separation.
This, however, should not discourage use of the MPS method, which indeed does not rely on the particular fall-off of correlations with distance but rather on the entanglement being finite between any two parts, e.g., defined by a cut at any given location in the chain.
The fact that the global mode mediates interactions between any such parts of the spin chain may not cause a problem, as long as their entanglement is kept finite.
Our study, in parts, aims to investigate how crucial are the resulting limitations on the Hilbert space dimension and the degree of excitation of the global system.

The natural way to represent our system is to decompose the system Hilbert space into a spin chain-part and the external, global part $\H = \H_\text{lattice} \tp \H_\text{ext}$.
The spin-chain space is a tensor product of $L$ copies of a $d$-dimensional space $\H_\text{lattice} = \H_\text{site}^{\tp L}$.
The system state can then be expanded as a spin chain-state for each basis-state of the global degree of freedom, e.g.
\begin{align}
 \ket\psi = \sum_{n=0}^{d_\text{ext} - 1} \ket{\psi_n} \tp \ket{n}, \label{eq:MPSexpansion}
\end{align}
where $\ket n$ is the $n$'th basis-state of the external degree of freedom and $\ket{\psi_n}$ could then be an MPS with tensors $A^{[i,n]}$ defined for each site $i$ and each global system basis state $n$.
If the external degree of freedom is a a spin of magnitude $S$, $d_\text{ext} = 2S + 1$, while for a harmonic oscillator a truncation would be necessary and $d_\text{ext}$ will then denote the cut-off excitation number.

By choosing $\ket{\psi_n}$ as MPS-states we can also rewrite the full state $\ket\psi$ as a block-diagonal MPS-state, i.e.
\begin{align}
 \ket\psi = \sum_{i_1,\ldots i_L, n} \trace(v^n A^{[1],i_1} \ldots A^{[L],i_L} ) \ket{n, i_1, \ldots i_L} \label{eq:CentralMPS}
\end{align}
where
\begin{align*}
 A^{[m],i_m} = \begin{pmatrix}
        A^{[m,1],i_m} & \\
	& A^{[m,2],i_m} & \\
	& & \ddots
       \end{pmatrix}
\end{align*}
and $v^n$ is a vector of zero matrices except the block corresponding to the $n$'th component of $A^{[1]}$ which is an identity.
Note that (\ref{eq:CentralMPS}) is equivalent to a standard MPS ansatz, only it is restricted to a certain block-diagonal structure, with a new effective $\tilde D = D d_{ext}$.
In practice, however, we do not need to restrict to this block-diagonal form.
By allowing general matrices we can have a more compact representation when, e.g. the two subsystems are in a product state.

For systems with a few degrees of freedom interacting with a spin chain the Hamiltonian will most naturally contain terms which are local in the chain and the external degrees of freedom, or which are simple products of such operators.
All such terms have compact matrix product operator expressions and the tensor networks arising in matrix-elements can be evaluated efficiently.
For example, an interaction term $V = A \sum_{i=1}^L X_i$ where $A$ acts on $\H_\text{ext}$ and $X_i$ acts on the $i$'th lattice site can be written as a matrix product operator (MPO) with tensors
\begin{align*}
 M^0 = (A)
\qquad
 M^1 = \begin{pmatrix}
        X_1 & I
       \end{pmatrix},
 M^i = \begin{pmatrix}
        I & 0 \\ X_i & I 
       \end{pmatrix},
 M^L = \begin{pmatrix}
        X_1 \\ I 
       \end{pmatrix}
\end{align*}
where $A$ is a $d_{ext}\times d_{ext}$ matrix, and the $X_i$ are $d\times d$ matrices.
These compact representations of operators are key to the application of the variational methods developed for matrix product states; both for eigenstates \cite{verstraete_matrix_2008} and time-evolution \cite{garcia-ripoll_time_2006}.

For unitary time-evolution we will use the Krylov propagation algorithm \cite{garcia-ripoll_time_2006} instead of the more common Suzuki-Trotter decomposition.
The Krylov method consists of constructing a basis for the subspace spanned by $\{\ket\psi, H\ket\psi, H^2\ket\psi, \ldots, H^N \ket\psi \}$.
By truncating at a finite power of the Hamiltonian we can get a good approximation to the matrix exponential of $H$ and thus perform time-evolution very efficiently within this subspace.
Once the state $\ket\psi$ has been propagated within the Krylov subspace, it is injected into the full Hilbert space again.

In order to perform these operations for an MPS we need to be able to construct the Krylov basis and inject the propagated vector into the Hilbert space using variational methods.
There is an efficient algorithm for finding an MPS of a given virtual dimension which best approximates an operator-MPS product, i.e. find $\ket\chi$ such that $\norm{\ket\chi - T\ket\psi}^2$ is smallest.
This works essentially as long as the expression $\braket{\chi|T|\psi}$ can be evaluated efficiently.
The orthogonal Krylov basis vectors are given by the Gram-Schmidt algorithm as
\begin{align*}
 \ket{j} = H\ket{j-1} - \sum_{i=0}^{j-1} \frac{\braket{i|H|j - 1}}{\braket{i|i}} \ket i,
\end{align*}
for $j=1, \ldots N - 1$ and $\ket{0} = \ket\psi$.
Since we can contract terms of the form $\braket{i|H|j}$ when $\ket i$, and $\ket j$ are matrix product states efficiently, we can also find an MPS which is closest to each of the Krylov basis vectors efficiently.
After performing the unitary time evolution within the Krylov subspace we have a state $U\ket\psi \approx \sum_{j=0}^{N-1} c_j \ket j$ for which the nearest MPS is found using the the same variational technique.

The ground state calculations presented in this manuscript were all performed with sweeps of local minimization of the effective Hamiltonian \cite{verstraete_matrix_2008} (using $D \sim 128$), whereas the time-evolution was calculated using the Krylov subspace technique described in \cite{garcia-ripoll_time_2006} (using $D \sim 512$).

\section{The Dicke-Ising model}

\begin{figure*}[htb]
 \includegraphics[width=0.5\columnwidth]{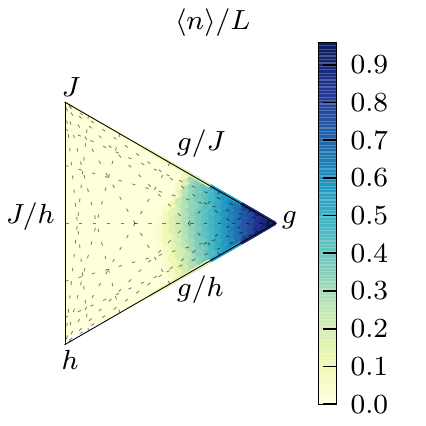}
 \includegraphics[width=0.5\columnwidth]{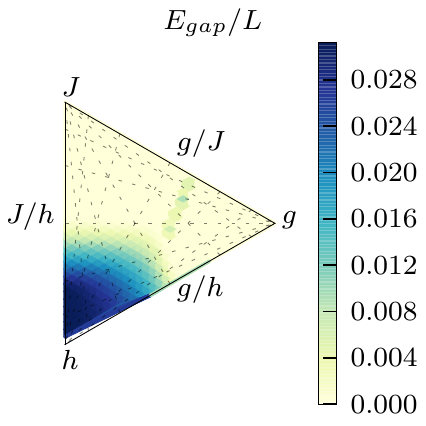}
 \includegraphics[width=0.5\columnwidth]{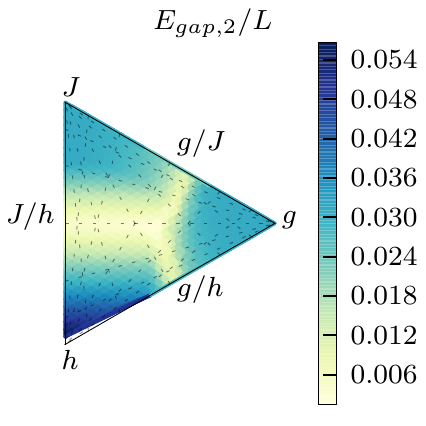}
 \includegraphics[width=0.5\columnwidth]{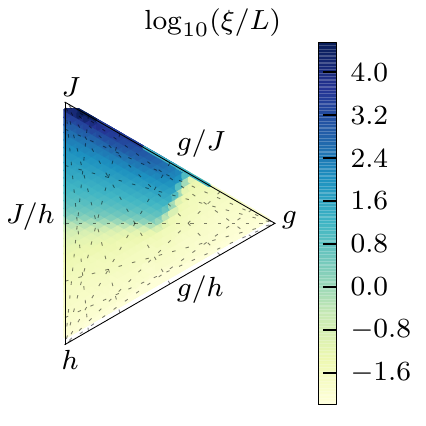}
 \caption{(Color online) From left to right: expected photon density $\braket{n} / L$, Energy gap to the excited state, Energy gap to the second excited state and the correlation-length for $\sigma^y$.
Values are for 32 spins with $\omega = 1$, and $h + J + g = 1$.}
 \label{fig:pd}
\end{figure*}

We describe our interacting spin chain and the global oscillator by the Hamiltonian,
\begin{align}
 H = \omega a^\dagger a - h \sum_{j=1}^L \sigma_z^j - J \sum_{j=1}^{L-1} \sigma_y^i \sigma_y^{i+1} + \frac{g}{\sqrt{L}} \sum_{j=1}^L \sigma_x^j (a + a^\dagger), \label{eq:DickeIsingH}
\end{align}
where $\sigma^j_\xi$, $\xi=x,y,z$ denote Pauli spin-1/2 operators for the $j^{th}$ spin, $\omega, a,a^\dagger$ denote the eigenfrequency, and the lowering and raising operators for the oscillator, $J$ is the Ising coupling constant, $g/\sqrt{L}$ is the Dicke coupling constant, normalized to the system size, and $h$ is a transverse bias (magnetic) field.
The goal of our calculations is to identify the phase diagram under variation of the control parameters $J, h, g$ in units where $\omega = 1$.

\begin{figure}[htb]
 \includegraphics[width=\columnwidth]{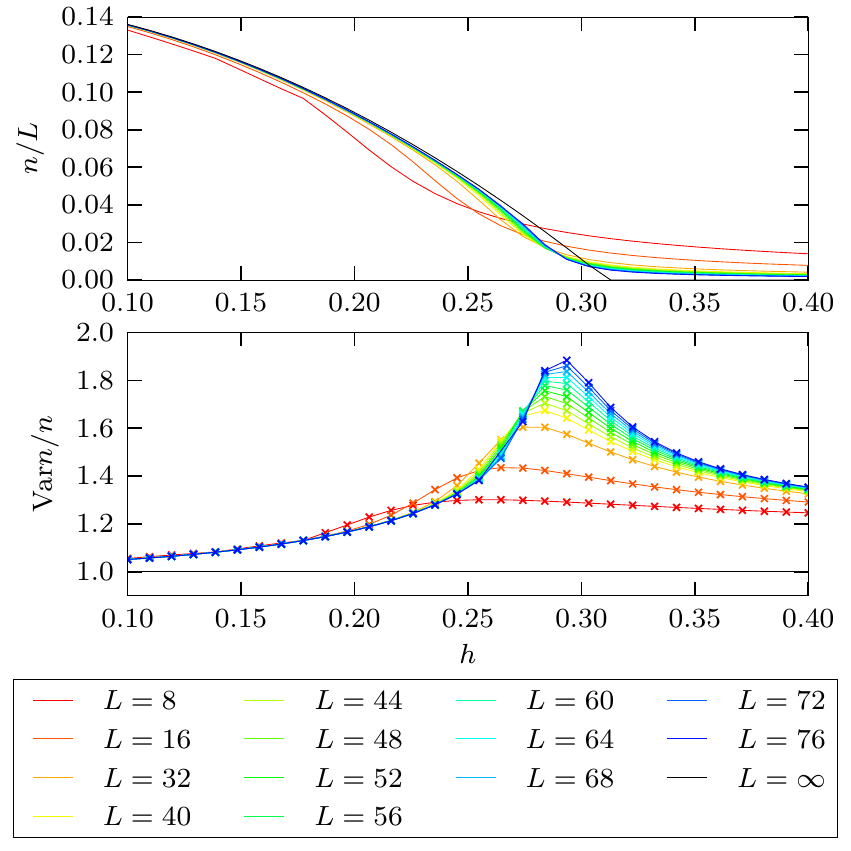}
\caption{(Color online) Photon number mean (upper panel) and variance (lower panel) as functions of $h$ for $\omega = 1$, $g = 0.4$, $J = 0.1$.
Note the peak in $\Var n / n$ indicating a super-poissonian distribution at the phase transition.
As a function of $L$ the peak height scales as $\Var n / n \propto L^{0.19}$ and the peak location scales as $\abs{h - h_c} \propto L^{-0.5}$ where $h_c$ is the location of the critical point.
}
\label{fig:slice2ndorder}
\end{figure}

\begin{figure}[htb]
 \includegraphics[width=\columnwidth]{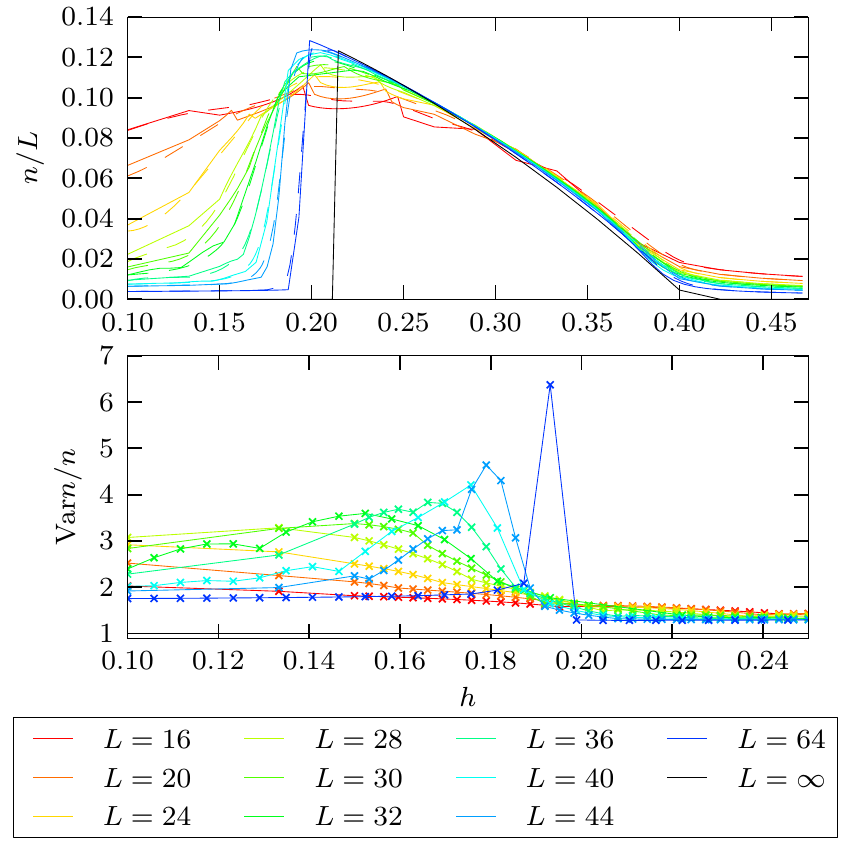}
\caption{(Color online) $\omega = 1$, $g = 0.49$, $J = 0.3$.
The upper panel shows the mean photon number as a function of $h$ near the first order transition.
The lower panel shows the photon number variance in the region from $h = 0.1$ to $h = 0.25$.
The dashed curve in the upper figure shows the mean photon number for a low, but finite, temperature. 
}
\label{fig:slice1storder}
\end{figure}

The analyses in \cite{gammelmark_phase_2011-1,lee_first-order_2004} suggest that in the thermodynamic limit $L\to\infty$ a mean field description of the oscillator mode should be valid.
The mean field approximation essentially implies that the oscillator-spin chain system state is factored into a coherent state $|\alpha\rangle$ and an Ising-interacting chain with an effective transverse magnetic field $h_\text{eff}^2 = h^2 + 4g^2 \Re(\alpha)^2$.
We expect that the MPS ansatz (\ref{eq:CentralMPS}) should work well in this limit, since it only requires tensor-dimension 1 on the link connecting the oscillator and the spin chain.

The Hamiltonian (\ref{eq:DickeIsingH}) has a discrete symmetry $P = \pi \tp \sigma_z^1 \ldots \sigma_z^L$ where $\pi^\dagger a \pi = -a$.
This symmetry is broken in the mean field description since $\pi \ket{\alpha} = \ket{-\alpha}$ and the free energy in the thermodynamic limit is invariant under the replacement $\alpha \to -\alpha$.
For a finite-size system the symmetry is not broken and the system ground state is either a $+1$ or $-1$ eigenstate of $P$.
Since the mean field theory predicts $\alpha$ proportional to $\sqrt{L}$, and $\braket{\alpha|-\alpha} = \exp(-2\abs{\alpha}^2)$, where matrix elements between $\ket\alpha$ and $\ket{-\alpha}$ of $a^\dagger a$ and $a + a^\dagger$ will be exponentially suppressed with increasing $L$,  it becomes exceedingly difficult, even for moderate systems sizes, to numerically identify states with the proper symmetry of the Hamiltonian without explicit attention to symmetry.

We have used the MPS method to calculate the properties of the lowest eigenstates of the Hamiltonian (\ref{eq:DickeIsingH}) for $L = 32$ spins.
In Figure \ref{fig:pd} we have plotted the values of different physical quantities in four simplex plots.
The simplex plots restricts the four control parameters by fixing one of them, e.g., $\omega = 1$, and by imposing a linear constraint among the others, such as $h + g + J = 1$ (assuming also that $h,g,J \geq 0$).
The parameter dependence of the given physical quantities are then shown by color (hue) in the resulting 2D plot.
In any corner of the plots, one parameter attains its maximum and the others vanish, while along the opposing edge the same parameter vanishes while the other two vary under the constraint that their sum is fixed.
Along the $h$-$J$-edge (the vertical edge in Figure \ref{fig:pd}) the coupling between the oscillator and the spins vanishes, and the model reduces to the conventional Ising model, while for the lower ($h$-$g$) edge, the model has no spin-spin interactions and therefore corresponds to the Dicke phase transition model.

The first panel in Figure \ref{fig:pd} shows the expected excitation of the oscillator divided by the number of spins, and we observe a strong similarity with the results of the low temperature mean field theory (see Figure 1 in \cite{gammelmark_phase_2011-1}).
Due to the finite size of the system, phase transitions will be smoothed, but still, we do recognize the characteristics of the Dicke second order phase transition from the normal to the super-radiant state for low spin-spin interaction strength.
The similarly smoothed characteristics of the Ising-transition can also be seen in the fourth panel.
The spins change from pointing parallel to the $z$-axis to having no net magnetization but strong $y-y$-correlations, $\langle \sigma_y^i \sigma_y^j \rangle \simeq e^{-|i-j|/\xi}$.
Finally, the first order transition predicted by \cite{lee_first-order_2004, gammelmark_phase_2011-1} can also be seen along the upper edge in the simplex.

The excited state energy gaps (panels two and three in Figure \ref{fig:pd}) in our finite system calculations also reflect the significant difference between the first and second order transitions.
The second order transition occurs from the $J < h$ region of the phase diagram and the system therefore enters from a unique ground state with a finite total magnetization in the $z$-direction.
During the second order transition the two lowest eigenstates transform from spins pointing in opposite $z$-directions to the two degenerate mean-field states.

In Figure \ref{fig:slice2ndorder} we investigate the effects of the finite size of the system in more detail where the mean excitation of the oscillator, normalized to the number of spins and its variance are plotted as functions of external magnetic field $h$ with $\omega = 1$, $g = 0.4$ and $J = 0.1$ for a variety of system sizes.
According to the mean field theory, a second order phase transition into the super-radiant state occurs at $h \approx 0.312$ with the $q = (a + a^\dagger)/\sqrt{2}$-field quadrature as order parameter.
This is indicated with a full  black line in the upper panel.

For small system sizes the system gradually changes from supporting a few to a larger number of oscillator excitation quanta per particle and the MPS results gradually approach the ones of the mean field theory.
The degree of excitation quickly approaches the mean field value as the system size increases and already for $L = 32$ spins the system behaves similarly to the mean field theory.
An important difference between the mean field theory and the numerical calculations is the normalized variance $\Var n / n$ shown in the lower panel.
The mean field theory predicts a Poisson distribution for the photon number for all parameter values implying $\Var n / n = 1$; in the numerical calculations which go beyond the mean field description, there is a substantial deviation from this result near the second order transition where we clearly see a peak in the normalized excitation variance scaling as $\Var n / n \propto L^{0.19}$  for this set of parameters.

This peak is most likely due to the oscillator ground state gradually splitting into two coherent components giving rise to a widening of the excitation number distribution.
In the normal phase the cavity is close to a vacuum state and develops first into a weakly excited state consisting primarily of the symmetric combination $\ket{\alpha} + \ket{-\alpha}$, with an excitation number variance which is higher than the individual coherent-state components.
As the system enters the super-radiant state the field components separate and become correlated with orthogonal states of the spin chain.
The field state then becomes equivalent to an even statistical mixture of the two symmetry-broken states with conventional Poisson statistics.

In Figure \ref{fig:slice1storder} we have plotted the photon number density and $\Var n / n$ near a first-order transition.
Near the phase transition one can discern oscillations in the photon number due to the ground state and first excited state alternating between two close-lying levels with different symmetry.
At finite temperature these oscillations are washed out as can be seen by following the dashed curves in Figure \ref{fig:slice1storder}.

Accompanying the oscillations in the mean photon number, significant deviations from the Poissonian statistics expected from the mean-field theory, can be seen in the lower part of Figure \ref{fig:slice1storder}.
A striking feature in this parameter regime is the significant effect of the finite size of the system.
Here, $L > 40$ is needed in order to see a sharp drop in photon number below the critical point, and the photon number variance is very far from the mean-field result for even larger systems.
In addition, the location of the first-order transition predicted by mean-field theory is only approached very slowly as a function of the system size.

\section{Time evolution during homodyne detection}

\begin{figure}
 \includegraphics[width=\columnwidth]{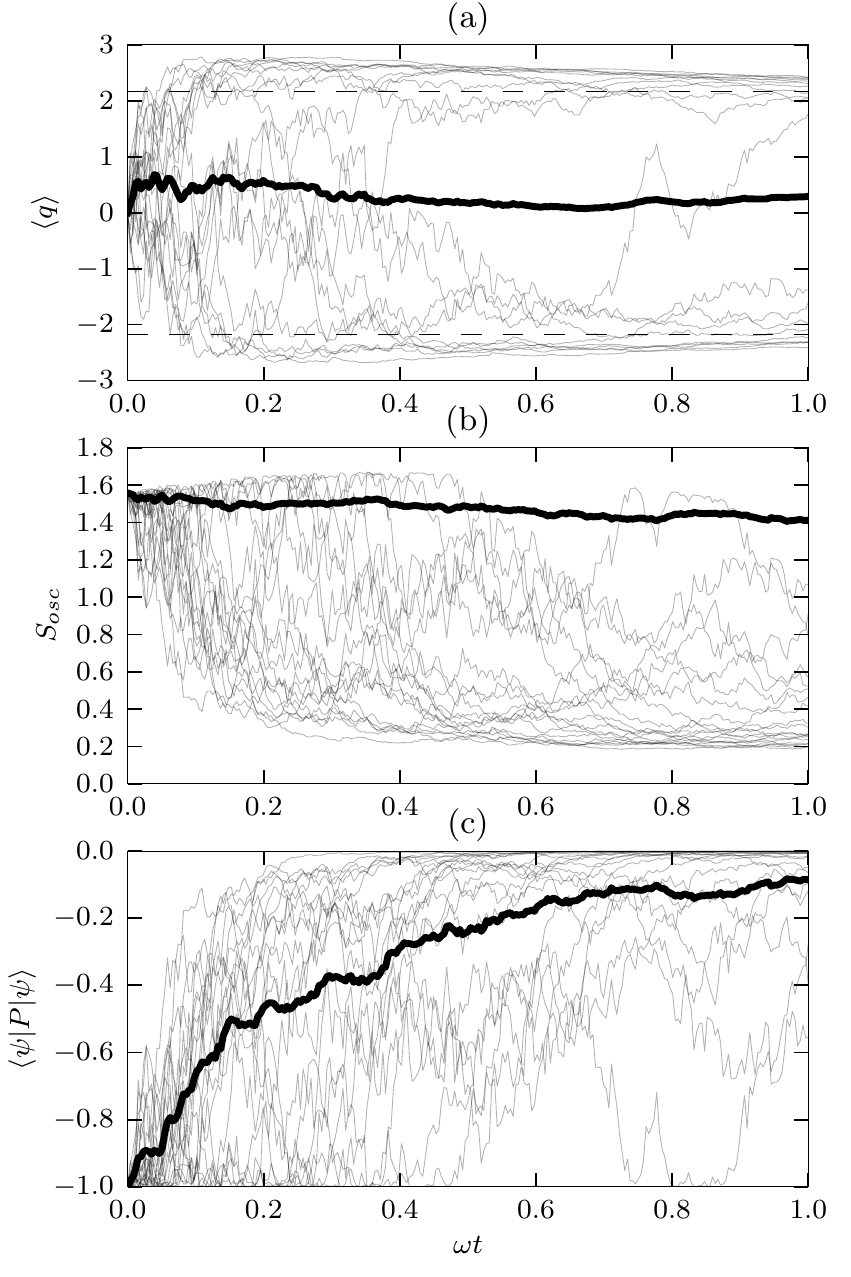}
\caption{(Color online) Time evolution of system being monitored by homodyne detection.
(a) The conditioned expectation value of the $a + a^\dagger$ quadrature of the cavity field.
The dashed line are the values of $q$ for the two symmetry-broken states in the thermodynamic limit.
(b) The thick line is the von Neuman entropy of the average time-evolved oscillator state, and the thin lines are the entropy of the oscillator state in a few sample trajectories.
(c) The expectation-value of the symmetry operator $P$. 
}
\label{fig:timeevolution}
\end{figure}

While the Hamiltonian commutes with the symmetry operator $P$ and the energy eigenstates can be chosen as eigenstates of $P$, the symmetry breaking ansatz associated with either one of the coherent states is a significant simplification of the problem and it makes accurate predictions in the macroscopic limit.
The spontaneous symmetry breaking ansatz may be justified by the smallness of any perturbation needed to make the symmetry broken state the actual ground state of the system, but it may also be justified by the coupling and entanglement of the system with its environment.
A measurement taking place within the environment gives a random outcome by the laws of quantum mechanics, and this randomness may, indeed, be sufficient to break the symmetry of the quantum state.

In our case, it is natural to imagine that the cavity is lossy, and thus the field may leak out of the cavity and become subject to a phase sensitive measurement probing.
We consider a situation where the emitted cavity field is measured with a homodyne detection scheme.
The system then evolves according to the stochastic Schr\"odinger equation \cite{wiseman_quantum_1996,goetsch_linear_1994,jacobs_straightforward_2006}
\begin{multline*}
 d\ket\psi = \left[-i H dt - \frac{\kappa}{2}\left( a^\dagger a - \braket{a + a^\dagger} a + \frac{1}{4}\braket{a + a^\dagger}^2 \right) dt \right. \\
  \left. + \sqrt{\kappa}(a - \frac{1}{2}\braket{a + a^\dagger}) dW \right] \ket\psi
\end{multline*}
where $(a + a^\dagger)/\sqrt{2}$ is the cavity quadrature component being measured by the homodyne detector, $dW$ is a Wiener increment and $\kappa$ is the cavity loss rate.

To simulate this equation, we treat the unitary and stochastic dynamics using a split-step approach as in \cite{gammelmark_simulating_2010}.
The system is first propagated within a Krylov subspace generated by $H$ for a time $\Delta t$ and the stochastic part can be propagated by selecting a normally distributed random number $\Delta y$ with variance $\Delta t$, and mean $\sqrt{\kappa}\braket{a + a^\dagger} dt$ and update the MPS-state according to the action of the operator
\begin{align*}
 \Omega(\Delta y) = \id - \frac{\kappa}{2} a^\dagger a dt  + \sqrt{\kappa}a \Delta y
\end{align*}
followed by renormalization \cite{wiseman_quantum_1996}.
This mimics the measurement back action and samples the distribution of observed signals $\Delta y(t)$.
The measurement back action acts locally on the system, and hence the determination of the resulting MPS state is efficient \cite{gammelmark_simulating_2010}.

In Figure \ref{fig:timeevolution} we show a small ensemble of simulated trajectories for a system of 32 spins, $\omega = 1$, $h = 0.2$, $J = 0.3$ and $g = 0.49$ and $\kappa = 0.5$.
In Figure \ref{fig:timeevolution}(a) the thin lines show the conditioned time-evolution of the expected value of the cavity quadrature $q = (a + a^\dagger)/\sqrt{2}$ along with the value estimated from the two symmetry-broken thermodynamic states (dashed line). 
The $q$-quadrature starts out at zero, but quickly stabilizes to a value of about $\pm 2.1$.
This is consistent with the interpretation that by probing the cavity field the observer is able to distinguish the two symmetry-broken states and thereby drives the system to the state consistent with the observation.

The time-scale on which the states are distinguished corresponds nicely with the time-scale for the decay of a mesoscopic superposition state \cite{PhysRevLett.77.4887}.
This time-scale is given by $\tau \approx 1/\kappa \abs{\Delta \alpha}^2$, where $\Delta\alpha$ is the seperation between the two coherent components.
For $\kappa = 0.50\omega$ and $\abs{\Delta\alpha} \approx 4.2$ we estimate $\omega\tau \approx 0.11$, which is consistent with the time-scale observed in Figure \ref{fig:timeevolution}.
The thick line in Figure \ref{fig:timeevolution}(a) is the ensemble average of the field quadratures.
This ensemble average, corresponding to an unobserved, dissipative evolution, remains close to zero, since, on \emph{average}, the observer will detect the two symmetry-broken states an equal number of times.

The second panel (b) in Figure \ref{fig:timeevolution} shows the von Neumann entropy $S(\rho) = -\trace(\rho \log_2 \rho)$ for the oscillator state.
In general, the entropy for the individual trajectories are decreasing towards zero, again consistent with the cavity approaching a pure state.
The average (dissipative) evolution, however, does not have decreasing entropy.
This is another signature of the fact that, for specific trajectories the observers knowledge is sufficient to distinguish the two symmetry-broken states.

If the initial state was a pure superposition of the two many-body symmetry-broken states, the oscillator would be in an equal binary mixture of the two coherent states $\ket\alpha$ and $\ket{-\alpha}$ and we would expect $S = 1$.
However, in the case shown here, $S(0) \approx 1.6$ indicating that the initial state is not well-described by two orthogonal symmetry-broken states.
This is not particularly surprising in this case, since the system is of moderate size ($L = 32$) and the chosen initial state is close to the first order phase transition shown in Figure \ref{fig:slice1storder}.

Note, that $S$ is also a measure of the number of states in a Schmidt-decomposition of the state into oscillator- and spin-chain sub systems.
Since the oscillator entropy is small for an accurate representation of the state (the ground state energy is accurate to 5 decimal places) we conclude that, in this case, the non-local interactions do not present a major problem for the MPS techniques.

The third panel (c) in Figure \ref{fig:timeevolution} shows the expectation value of the symmetry-operator $P = \pi \tp \sigma_z^1 \ldots \tp \sigma_z^L$.
The figure clearly shows that the initial ground state is anti-symmetric and that the symmetry is lost in the subsequent time-evolution where one of the symmetry-broken states is selected.

\section{Conclusion}

In this paper we have applied matrix product state-methods to investigate ground-state and dynamic properties of the Dicke-Ising model.
The Dicke-Ising model not only has interactions between spins, but also a global cavity mode, interacting uniformly with all the spins.
This implies that entanglement can exist over arbitrary distances in the system which, in turn, can be a challenge for the numerical treatment of the problem.
We have shown that the matrix product state method with a Krylov subspace expansion of the propagator permits effective numerical treatment of the problem, and we have identified the phase transitions in the model predicted by mean-field theory. 

Near the second order phase-transition the mean-field solution for the average photon number is readily reproduced, while the photon number variance deviates significantly from the mean-field results.
The increased variance is ascribed to the formation of a superposition of two coherent states near the phase transition in contrast to the even statistical mixture of coherent states predicted by mean-field theory.
Near the first order phase transition there are significant and non-trivial finite-size effects causing a weak thermal-like excitation of the cavity mode below the phase transition, which approaches the vacuum state only for very long spin chains.

The interplay of quantum mechanical measurements and many-body physics is an interesting combination of thermodynamics and information theory and to investigate the phase transition dynamics further, we have simulated the conditioned stochastic dynamics of the Dicke-Ising system when it is subject to homodyne detection of the field leaking from the cavity.

Quantum mechanical measurements serve as an attractive model for understanding the symmetry-breaking mechanism.
Using this approach we can not only explain why the symmetry-breaking happens, but also the associated time-scale.
As clearly shown by our simulations, the measurement back action causes a dynamic symmetry breaking, as the observer gradually distinguishes between the cavity output from the two candidate (mean-field) symmetry-broken states.

The authors acknowledge support from the EU integrated project AQUTE.

\bibliography{litterature}

\end{document}